\documentclass[12pt]{article}

\usepackage{graphicx}
\usepackage{psfrag}
\usepackage{flafter}

 \newcommand {\nc}{\newcommand}

 \nc{\be}{\begin{equation}}

 \nc{\ee}{\end{equation}}

 \nc{\bea}{\begin{eqnarray}}

 \nc{\eea}{\end{eqnarray}}

 \def\complex{{\cal C}}

 \def\integer{{\cal Z}}

 \def\natural{{\cal N}}

 \def\unit{{\bf 1}}

 \def\tld{\tilde}

 \def\bra{\langle}

 \def\ket{\rangle}

 \title{Quiver Matrix Mechanics for IIB String Theory (I):\\
 Wrapping Membranes and Emergent Dimension}
 \author{
  Jian Dai\thanks{\tt jdai@physics.utah.edu}$\;\,$
  and Yong-Shi Wu\thanks{\tt wu@physics.utah.edu}\\
  Department of Physics, University of Utah\\
  Salt Lake City, Utah 84112}

  \date{\today}

\begin{document}


  \maketitle

  \begin{abstract}

   In this paper we present a discrete, non-perturbative
   formulation for type IIB string theory. Being a
   supersymmetric quiver matrix mechanics model in the
   framework of M(atrix) theory, it is a generalization of
   our previous proposal of compactification via orbifolding
   for deconstructed IIA strings. In the continuum limit,
   our matrix mechanics becomes a $(2+1)$-dimensional Yang-Mills
   theory with 16 supercharges. At the discrete level, we
   are able to construct explicitly the solitonic states
   that correspond to membranes wrapping on the compactified
   torus in target space. These states have a manifestly
   $SL(2,\integer)$-invariant spectrum with correct membrane
   tension, and give rise to an emergent flat dimension
   when the compactified torus shrinks to vanishing size.

  \end{abstract}
  \vspace{0.25in}


  \section{Introduction}
  \label{SEC1}

   In the string/M-theory duality web, both the IIA/M and IIB/M
   duality conjectures involve the emergence of a new spatial
   dimension in a certain limit. M-theory, according to our
   current understanding, is the 11-dimensional dual of the
   strongly coupled 10-dimensional IIA string theory \cite{WIT95}.
   Here a new dimension emerges, because the D-particles,
   carrying RR charges, can be viewed as the Kaluza-Klein
   momentum modes along a circle in the ``hidden eleventh''
   dimension. The size of the circle depends on the IIA string
   coupling in such a way that when the string coupling tends
   to infinity, the size of the circle also becomes infinite.
   In other words, IIA string theory should be dual to M-theory
   compactified on a circle, and the hidden dimension opens up
   as a new flat dimension in the strong-coupling limit.

   For IIB/M duality, the story is obviously more complicated
   \cite{SCH,SCH1,AS}. IIB string theory is conjectured to
   be dual to M-theory compactified on a 2-torus. The IIB
   string coupling depends on the area of the 2-torus, in
   such a way that the weak coupled IIB string theory
   should be recovered when the area of the hidden 2-torus
   shrinks to zero. A great advantage of this scenario is
   that the mysterious $SL(2,\integer)$ duality in IIB
   string theory can be interpreted as the geometric,
   modular transformation of the hidden 2-torus. But wait,
   there is apparently a mismatch of dimensionality here:
   with the compactified two dimensions diminished in M
   theory, it seems that only eight spatial dimensions remain,
   but IIB string theory lives in nine, not eight, dimensional
   space. The clever solution \cite{SCH} to this puzzle is
   that in M theory there are topological soliton-like states,
   corresponding to wrapping a membrane around the compactified
   2-torus. Because of the conjectured $SL(2,\integer)$
   invariance, the energy of the wrapping membrane should only
   depend on its wrapping number $w$. With the torus shrinking
   to zero, a tower of the wrapping membrane states labeled by
   $w$ gives rise to the momentum states in a new, flat, emergent
   dimension.

   The picture looks perfect. But can we really formulate it
   in a mathematical formalism with above-mentioned ideas
   demonstrated explicitly? Certainly this needs a non-perturbative
   formulation of M-theory. Though a fully Lorentz invariant
   formulation of M-theory is not available yet, fortunately we
   have had a promising candidate (or conjecture), namely the
   BFSS M(atrix) Theory \cite{BFSS}, in a particular kinematical
   limit, the infinite momentum frame (IMF), or in the discrete
   light-cone quantization (DLCQ) \cite{Sus97}. It was shown
   that M(atrix) Theory compactified on an $n$-torus (for $n\le
   3$) is just an $(n+1)$-dimensional supersymmetric Yang-Mills
   (SYM) theory \cite{Tay}. Facilitated by this fact, the IIA/M
   duality was verified soon after BFSS conjecture. Namely
   M(atrix) Theory compactified on a circle was explicitly
   shown, via the so-called ``9-11 flip'', to give rise to a
   description of (second quantized) IIA strings \cite{DVV}.
   As to IIB/M duality, again the story was not so simple
   as IIA/M duality. The statement that M(atrix) theory
   compactified on a 2-torus (with one side in the
   longitudinal direction) is dual to IIB string
   theory compactified on a circle was checked \cite{HW}
   for the action of $D$-string \cite{SCH2} and the energy
   of $(p,q)$-strings etc. An emergent dimension with
   full $O(8)$ invariance together with seven ordinary
   dimensions were argued to come about in the
   $(2+1)$-dimensional SYM that results from compactifying
   M(atrix) Theory on a transverse torus \cite{S,SS,FHRS}.
   These arguments invoked the conjectured $SL(2,\integer)$
   duality of $(3+1)$-dimensional SYM. However, the explicit
   construction of wrapping membrane states and a proof of
   the $SL(2,\integer)$ invariance of these states, which
   plays a pivotal role in the IIB/M duality, are still
   lacking.

   Here let us try to scrutinize the origin of the difficulty
   for constructing a wrapping membrane state in M(atrix)
   theory. First recall that in the usual membrane theory,
   a nonlinear sigma model with continuous world-volume,
   a membrane state wrapping on a compactified 2-torus is
   formulated by
   \bea
   \label{WDM}
   \nonumber
    x^1(p,q)&=&R_1(mp+nq)+(\mbox{oscillator modes}),\\
    x^2(p,q)&=&R_2(sp+tq)+(\mbox{oscillator modes})
   \eea
   where $p$ and $q$ are two affine parameters of the
   membrane, two independent
   cycles on the compactified 2-torus have $R_1$, $R_2$ as their sizes and
   $x_1$, $x_2$ their affine parameters, $m,n,s,t$ four
   integers characterizing {\em linearly} how the membrane wraps on the
   torus. However, in M(atrix) Theory the target space
   coordinates are lifted to matrices, and the basis of the
   membrane world-volume functions, $e^{ip}$ and $e^{iq}$,
   are transcribed to noncommutative, (so-called) clock and
   shift matrices $U$ and $V$ of finite rank. Moreover, when
   the theory is compactified on a torus, the quotient
   conditions for compactification require the matrix coordinates of the target
   space to become covariant derivatives on the {\em dual
   torus}. Thus, the difficulty in constructing the matrix
   analog of the wrapping membrane states in Eq.~(\ref{WDM})
   lies in how to extract the affine parameters $p,q$ from
   their finite matrix counterparts $U$ and $V$, as well as
   to keep the linear relation containing the information of wrapping
   in Eq.~(\ref{WDM}).
   In a word, the problem is a mismatch between the linear coordinates of the target
   torus and the nonlinear coordinates of the membrane degrees of freedom.
   To our knowledge, no explicit construction to overcome this mismatch
   is known in the literature. The main goal of this paper
   is to fill this gap, namely we want to develop a M(atrix)
   Theory description of IIB string theory, which makes IIB/M
   duality and $SL(2,\integer)$-duality more accessible to
   analytic treatments.

   Our strategy to change the mismatch to be a match is
   to simply let both sides to be nonlinear; technically,
   we {\em deconstruct} the $(2+1)$-dimensional
   SYM that results from compactifying M(atrix) Theory on a torus
   and then, within the resulting framework of quiver matrix
   mechanics, to construct the matrix membrane wrapping states.
   In our previous paper \cite{DW}, we have been able to
   deconstruct the Matrix String Theory for type IIA strings,
   which is a $(1+1)$-dimensional SYM. Here we generalize
   this deconstruction to type IIB strings. The basic
   idea is to achieve compactification by orbifolding
   the M(atrix) Theory, like in Ref.~\cite{DW}. There,
   to compactify the theory on a circle, we took the
   orbifolding (or quotient) group to be $\integer_N$;
   but here to compactify the theory on a 2-torus we
   need to take a quotient with the group $\integer_N
   \otimes \integer_N$ (or shortly $\integer_N^2$). This
   leads to a supersymmetric quiver matrix mechanics with
   a product gauge group and bi-fundamental matter. By
   assigning non-zero vacuum expectation values (VEV) to
   bi-fundamental scalars, the quiver theory looks like
   a theory on a planar triangular lattice, whose continuum
   limit gives rise to the desired $(2+1)$-dimensional SYM.
   From the point of view of D-particles in target space,
   this is nothing but ``compactifying via orbifolding",
   or ``deconstruction of compactification''. (The use
   of the literary term and analytic technique of
   ``deconstruction" in high-energy physics was advocated
   in \cite{ACG}. Similar applications of the deconstruction
   method can also be found in \cite{RS} \cite{ACKKM}\cite{MRV}
   \cite{Kaplan}\cite{CEGK}\cite{KO}.) It is in this approach
   we have been able to accomplish a down-to-earth construction
   of wrapping membrane states in M(atrix) Theory, a modest
   step to better understand IIB/M duality.

   In this paper we will carry out our deconstruction for
   the simplest case with a right triangular lattice, and
   concentrate on the construction of the membrane states.
   The general case with a slanted triangular lattice and a
   detailed study of $SL(2,\integer)$ duality are left to the
   subsequent paper(s). We organize this paper as follows. In
   Sec.~\ref{SEC2}, we will give a brief review of deconstruction
   of the Matrix String Theory, to stipulate our notations and
   demonstrate the idea to approximate a compactification by
   an orbifolding sequence (in the so-called theory space).
   Then in Sec.~\ref{SEC3}, extending this approach, we
   achieve the deconstruction of the $(2+1)$-dimensional
   SYM that describes IIB strings by means of orbifolding
   M(atrix) Theory by $\integer_N\otimes \integer_N$,
   resulting in a supersymmetric quiver matrix mechanics.
   Metric aspects of this de(construction) after assigning
   VEV's is explored in Sec.~\ref{SEC4}; then the desired
   SYM is shown to emerge in the continuum limit with
   $N\to\infty$ upon choosing the simplest (right)
   triangular lattice. Sec.~\ref{SEC5} is devoted to the
   central issue of this paper, namely explicit construction
   of topologically non-trivial states in quiver mechanics
   that in the continuum limit give rise to the membranes
   wrapping on the compactified torus. The energy spectrum
   of these states is shown to depend only on the
   $SL(2,\integer)$-invariant wrapping number $w$. This
   provides an explicit verification of $SL(2,\integer)$
   invariance, as well as the correct membrane tension in
   our discrete approach. Moreover, the spectrum of these
   states is such that it can be identified as the spectrum
   of momentum states in an emergent flat transverse dimension
   when the size of the compactified torus shrinks to zero.
   Finally in Sec.~\ref{SEC6} discussions and perspectives
   are presented.

  \section{A Brief Review: Matrix String (De)Construction}
  \label{SEC2}

   This preparative section serves two purposes: to establish
   our notations and to review briefly the physical aspects of
   our previous (de)construction of the Matrix String Theory.

   \subsection{Preliminaries and Notations}

   By now it is well-known that at strong couplings
   of IIA string theory, a new spatial dimension
   emerges so that the 10-dimensional IIA theory is dual
   to an 11-dimensional theory, dubbed the name M-theory
   \cite{WIT95}. In accordance with the M(atrix)
   Theory conjecture \cite{BFSS} and as shown by Seiberg
   \cite{SEI} and Sen \cite{SEN}, the same dynamics that
   governs $N$ low-energy D-particles in a ten-dimensional
   Minkowski space-time actually captures all the
   information of M-theory in the discrete light-cone
   quantization. If we label the spatial coordinates
   by $y^1,y^2,\ldots,y^9$, then the basic idea of
   M(atrix) Theory is that, to incorporate open
   strings stretched between $N$ D-particles, one has
   to lift the D-particle coordinates to $N$-by-$N$
   matrices, $Y^I$ $(I=1,\cdots 9)$, as the basic
   dynamical variables.

   Formally the Matrix Theory {\it a la} BFSS can be
   formulated as a dimensional reduction of $d=9+1$,
   $\natural=1$ supersymmetric Yang-Mills theory:
   \[
    S = \int d^{10}x \{
     -{1\over {4g^2}} Tr F_{MN} F^{MN}
     -{i\over {2g^2}} Tr \lambda^T \Gamma^0\Gamma^M D_M\lambda
    \}
   \]
   where the Majorana-Weyl spinor $\lambda$, whose components
   will be labeled as $\lambda^{s_0s_1s_2s_3}$ ($s_a=1,2$,
   $a=0,1,2,3$), satisfy the spinor constraint equations
   $\lambda=-\Gamma^{11}\lambda$, $\lambda^\ast = \lambda$.
   The dimensional reduction follows the procedure below:
   $\int d^{10}x \rightarrow \int dt V_9$;
   $\tld{g}^2 := g^2/V_9$; then $A^I=:-X^I/\alpha^\prime$,
   $\tld{g}^2\alpha^{\prime 2} = g_s\alpha^{\prime 1/2}$,
   $\lambda^2/\tld{g}^2\rightarrow -\lambda^2$ and
   $X^I =: g_s^{1/3}\alpha^{\prime 1/2}Y^I$ where $I=1,\ldots, 9$.
   Note that in accordance with the IIA/M duality, the M-theory
   parameters are related to those of IIA string theory by
   $R=g_sl_s$, $l_p=g_s^{1/3}l_s$, in which $l_p$ is the
   eleven-dimensional Planck length and $R$ the radius of
   the hidden M-circle; In
   our convention, $\alpha^\prime =l_s^2$.
   Time variable and M-circle radius
   can be rescaled in units of Planck length to become
   dimensionless: $t=l_p \tau$, $R=R_{11}l_p$, respectively.

   After all these efforts, the BFSS action reads
   \be
   \label{BFSS}
    S = \int d\tau Tr \{
     {1\over 2R_{11}}[D_\tau, Y^I]^2 + \frac{R_{11}}{4}[Y^I,Y^J]^2
     -{i\over 2}\lambda^T[D_\tau,\lambda] + \frac{R_{11}}{2}
     \lambda^T \gamma^I [Y_I,\lambda]
     \},
   \ee
   in which eleven-dimensional Planck length is taken to be
   unity, $D_\tau= d/d\tau - i[Y^0,\cdot]$ is the covariant
   time derivative, and $Y^0$ is the gauge connection in the
   temporal direction. A representation of the gamma matrices
   $\gamma^I$ in Eq.~(\ref{BFSS}) is listed below for later
   uses:
   \bea
    \nonumber
     \gamma^0 := \unit \otimes \unit \otimes \unit \otimes \unit &,&
     \gamma^1 = \epsilon \otimes \epsilon \otimes \epsilon \otimes \epsilon, \\
    \nonumber
     \gamma^2 = \tau_1 \otimes \unit \otimes \epsilon \otimes \epsilon &,&
     \gamma^3 = \tau_3 \otimes \unit \otimes \epsilon \otimes \epsilon, \\
    \nonumber
     \gamma^4 = \epsilon \otimes \tau_1 \otimes \unit \otimes \epsilon &,&
     \gamma^5 = \epsilon \otimes \tau_3 \otimes \unit \otimes \epsilon, \\
    \nonumber
     \gamma^6 = \unit \otimes \epsilon \otimes \tau_1 \otimes \epsilon &,&
     \gamma^7 = \unit \otimes \epsilon \otimes \tau_3 \otimes \epsilon, \\
     \gamma^8 = \unit \otimes \unit \otimes \unit \otimes \tau_1 &,&
     \gamma^9 = \unit \otimes \unit \otimes \unit \otimes \tau_3,
   \eea
   where $\epsilon = i\tau_2$ and $\tau_a$ $(a=1,2,3)$ are Pauli
   matrices.

   In the following, the $N$-by-$N$ clock and shift matrices
   will be used frequently. We denote them by
   $U_N=diag(\omega_N,\omega_N^2,\ldots,\omega_N^N)$, where
   $\omega_N=e^{i2\pi/N}$, and
   \be\label{Shift}
    V_N:=
     \left(
      \begin{array}{ccccc}
       0&0&\cdots&0&1\\
       1&0&\cdots&0&0\\
       0&1&\cdots&0&0\\
       &&\ddots &&\\
       0&0&\cdots&1&0
      \end{array}
     \right).
   \ee
   Quite often we will just write $U$, $V$ when the rank
   $N$ is understood.

   \subsection{Compactification via Orbifolding}

   In this subsection we give a brief review of our
   previous work \cite{DW}, to show how the Matrix
   Theory compactification on a circle was achieved
   via orbifolding and how the worldsheet/target-space
   duality emerges in a natural manner.

   Consider a group of $K$ D-particles moving in
   a $\complex^2/\integer_N$ orbifold background;
   the orbifolding is performed via equivalence relations
   $z^1\sim e^{i2\pi/N}z^1$, $z^2\sim e^{-i2\pi/N}z^2$,
   where $z^1= (y^6+i y^7)/\sqrt{2}$, $z^2=(y^8+i y^9)/\sqrt{2}$.
   The complex coordinates $z^1$, $z^2$ can be parameterized
   in a polar coordinate form,
   $z^1=\rho_1 e^{i(\vartheta+\varphi)}/\sqrt{2}$,
   $z^2=\rho_2 e^{i(\vartheta-\varphi)}/\sqrt{2}$; the
   quotient conditions, in this parametrization, are
   simply $\varphi\sim\varphi+2\pi/N$. To consider $K$
   D-particles on the orbifold, we must lift the D-particle
   coordinate to $KN$-by-$KN$ matrices, for which we will
   use upper-cased letters $Z^1$, $Z^2$. To satisfy the
   orbifolding conditions these matrices are of the special
   form $z^a V_N$, where $z^a$ $(a=1,2)$ are in the form
   of $N\times N$ block-diagonal matrix
   with each block being a $K$-by-$K$ matrix and
   the unities in $V_N$ (see Eq.~(\ref{Shift}))
   understood as $\unit_K$.
   Besides orbifolding, we push the D-particles
   away from the singularity by assigning nonzero VEV to
   $z^1$ and $z^2$. Note that the moduli of the D-particle
   coordinates, $\bra z^1\ket = c_1/\sqrt{2}$,
   $\bra z^2 \ket = c_2/\sqrt{2}$, can be rotated to
   $\bra z^{1\prime}\ket = c_1^\prime /\sqrt{2}$,
   $c_1^\prime = \sqrt{|c_1|^2 + |c_2|^2}$
   with $\bra z^{2\prime}\ket = 0$, by an element of
   $SU(2)$ acting on the $\complex^2$. (The orbifolding
   results in a circular quiver diagram with $N$ sites.
   The $Z$-fields live on the links connecting neighboring
   sites, indicated by the presence of the block-matrix
   $V_N$.) So essentially only one modulus is needed to
   characterize the VEV.

   This modulus can be interpreted in two different
   pictures:
   \bea
   \label{TG}
   \mbox{I:}&&
    2\pi c_1^\prime = NL,\\
   \label{WS}
   \mbox{II:}&&
    N/c_1^\prime = 2\pi\Sigma.
   \eea
   On the one hand, $L$ is interpreted as the size of
   the circle on the orbifold in the target space, which is
   orthogonal to the radial direction and at the radius
   $c_1^\prime$. Viewed from the target space, D-particles
   live around this circle, so it can be viewed as the
   compactified circle, or the $M$-circle for IIA theory.
   On the other hand, $\Sigma$ in Eq. (\ref{TG}) can be viewed
   as the size of the circle that is dual to the target circle
   $L$. It is both amazing and amusing to see that with the
   fluctuations in $Z^1$ and $Z^2$ included, the orbifolded
   BFSS action can be equivalently written as an action on a
   one-dimensional lattice with lattice constant
   $a=2\pi\Sigma/N$.   Furthermore, in the limit $N\to\infty$,
   this quiver mechanics action approaches to the $d=1+1$ SYM
   (on the dual circle $\Sigma$) with 16 supercharges, known
   to describe the Matrix String theory, which was previously
   obtained by compactifying M(atrix) Theory on the circle
   $L$ directly. (Note that the so-called ``9-11 flip", that
   was necessary to give an appropriate interpretation for
   the compactified Matrix Theory on $L$, is not necessary
   at all in the orbifolding context).

\begin{figure}[hbtp]
    \centering
    \psfrag{Z}[][]{$c_1^\prime$}
    \psfrag{L}[][]{$L$}
    \psfrag{t}[][]{${2\pi\over N}$}
    \includegraphics[width=0.50\textwidth]{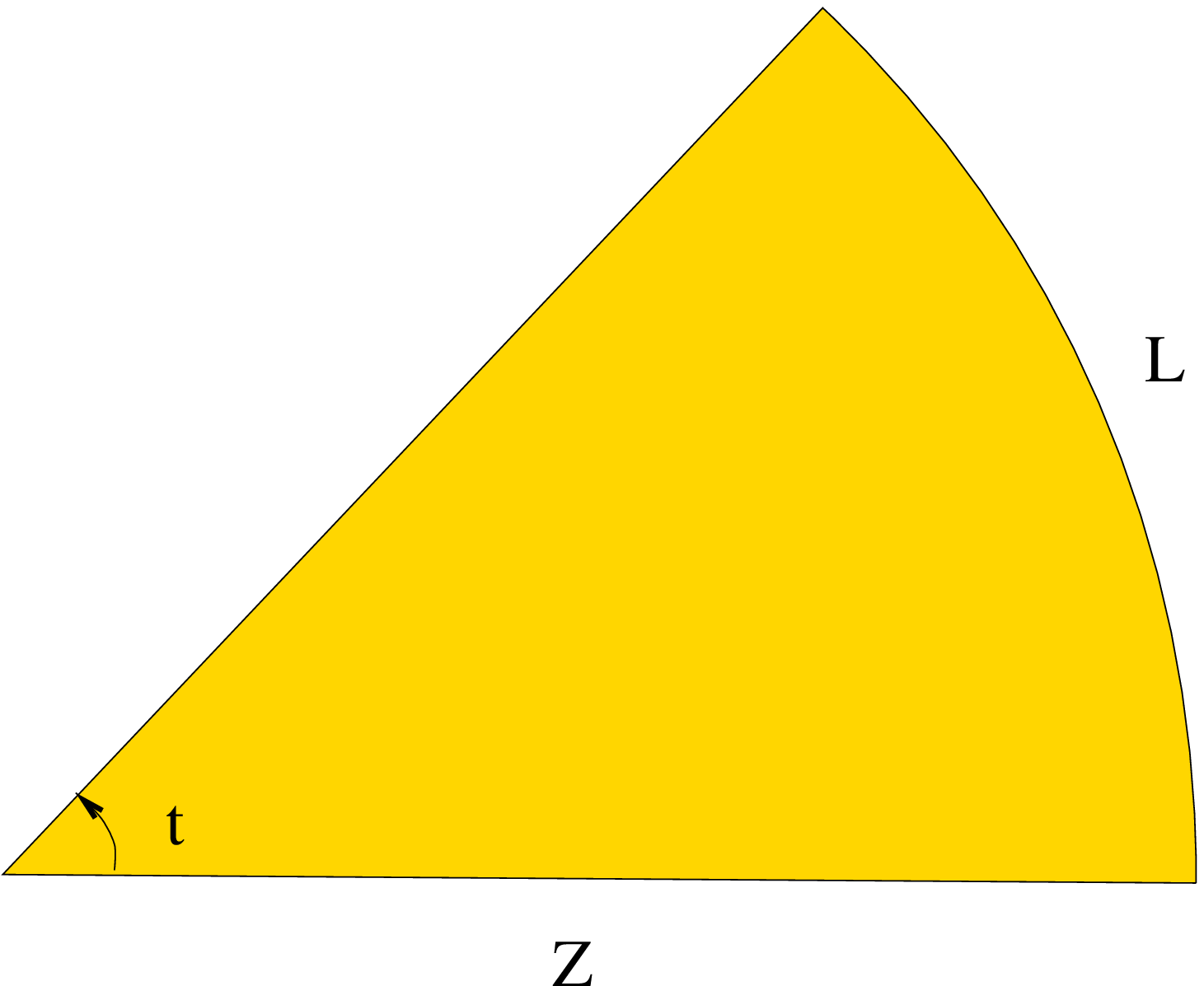}
    \psfrag{Z}[][]{$\Sigma$}
    \psfrag{L}[][]{${1\over c_1^\prime}$}
    \psfrag{t}[][]{${2\pi\over N}$}
    \includegraphics[width=0.40\textwidth]{Orb.eps}
    \caption{Worldsheet/target-space duality in
      $\complex^2/\integer_N$ Orbifold}
    \label{FIGORB}
   \end{figure}

   Thus, what we have done is actually to (de)construct
   the (type IIA) string worldsheet via a circular
   quiver diagram, and the relations (\ref{TG}) and
   (\ref{WS}) between the geometric parameters is a
   clear demonstration of the worldsheet/target-space
   duality: Namely Eq.~(\ref{TG}) is over target space
   orbifold, in which $c_1^\prime$ prescribes the
   location of the $M$-circle; Eq.~(\ref{WS}) is over
   worldsheet cycle, in which $c_1^\prime$ is the inverse
   of the lattice constant. The dual relation between the
   two interpretations and the compactification scale is
   shown in Figure \ref{FIGORB}.

\begin{figure}[hbtp]
    \centering
    \psfrag{N}[][]{$N$}
    \includegraphics[width=0.80\textwidth]{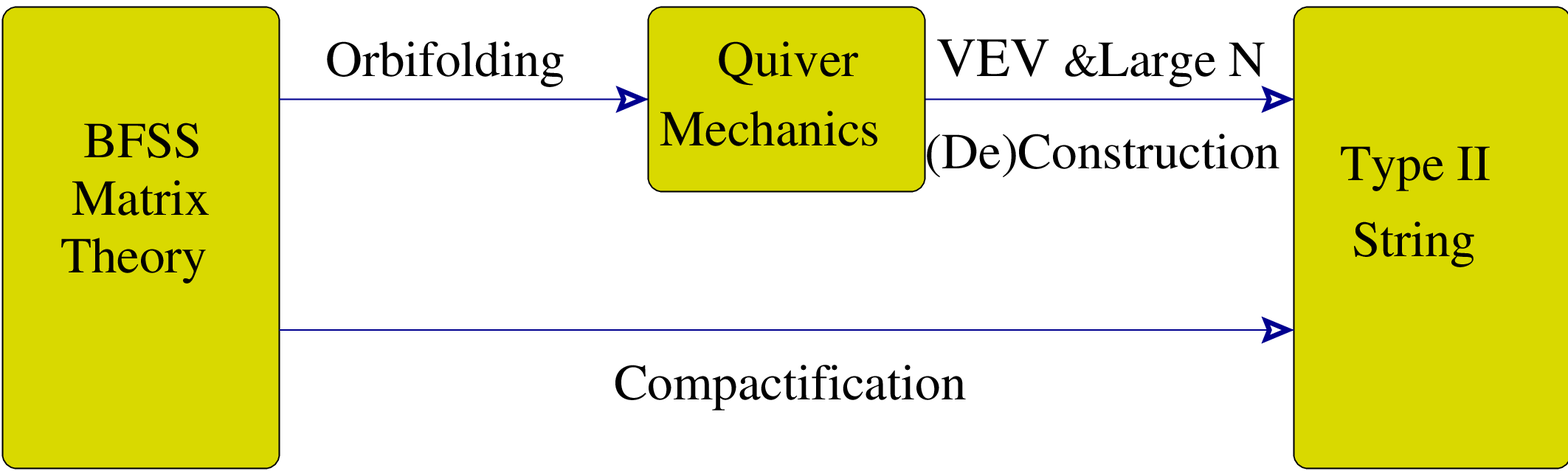}
\caption{Relation between Quiver Mechanics and Matrix
 Theory, Type II String Theory}
    \label{COMD}
   \end{figure}

   The methodology underlying the above deconstruction of
   IIA/M duality and the relationship between the Matrix
   Theory, type II string theory and our quiver quantum
   mechanics are demonstrated in a commuting diagram
   (Figure \ref{COMD}). In subsequent sections we will
   show that this diagram applies to the deconstruction
   of IIB strings as well. Additional complications from
   higher dimensions will emerge; and we will show how
   to deal with them in this and sequential paper.

  \section{$\integer_N\otimes\integer_N$ Orbifolded Matrix Theory}
  \label{SEC3}

   Now we proceed to make and justify our proposal that the
   $\integer_N\otimes \integer_N$ orbifolded M(atrix)
   Theory, being a supersymmetric quiver matrix mechanics,
   provides a promising candidate for the nonperturbative
   formulation of IIB string theory.

   Consider the orbifold $\complex^3/\integer_N \otimes
   \integer_N $, where the complex coordinates of
   $\complex^3$ are $z^a = (y^{2a+2}+iy^{2a+3})/\sqrt{2}$
   for $a=1,2,3$ and the quotient conditions for orbifolding are
   \bea
\mbox{I:} && z^1\sim e^{-i2\pi/N}z^1,
   z^2\sim e^{i2\pi/N}z^2, z^3\sim z^3;\\
\mbox{II:} && z^1\sim e^{-i2\pi/N}z^1,
   z^2\sim z^2, z^3\sim e^{i2\pi/N}z^3.
   \eea

   As did in the last section, we lift the D-particle
   coordinates to $KN^2$-by-$KN^2$ matrix variables
   $Z^a = (Y^{2a+2}+iY^{2a+3})/\sqrt{2}$, $(a=1,2,3)$,
   and subject them to the following orbifolding
   conditions:
    \be
\hat{U}_1^{\dag} Z^a \hat{U}_1 = \omega_N^{M_{45} - M_{67}} Z^a, \quad \hat{U}_2^{\dag} Z^a \hat{U}_2 =
\omega_N^{M_{45} - M_{89}} Z^a,
    \label{OCB}
   \ee
   in which $M_{IJ}$ are rotational generators
   on $IJ$-plane for vectors in nine-dimensional
   transverse space, and
   $\hat{U}_1:= \unit_K \otimes U_N \otimes \unit_N$,
   $\hat{U}_2: =\unit_K \otimes \unit_N \otimes U_N$
   embed the action of rotations into the gauge group,
   $U(KN^2)$, of M(atrix) Theory.

   Similarly introduce complexified fermionic coordinates
   for the Majorana-Weyl spinor $\lambda$ in M(atrix)
   Theory. For each real spinor index $s=1,2$ for $\lambda$,
   introduce complex spinor index $t=\pm$ through
   \[
    \left(
     \begin{array}{c}
\lambda^1\\
\lambda^2
     \end{array} \right)
    = T \left(
     \begin{array}{c}
\lambda^+\\
\lambda^-
\end{array} \right),
\qquad
    T = {1\over \sqrt{2}}
    \left(
     \begin{array}{cc}
      1&1\\ -i&i
     \end{array}
    \right).
   \]
   Therefore, $(\lambda^t)^\dag = \lambda^{-t}$. In addition,
   \[
    T^{\dag} \tau_1 T = - \tau_2,
    T^{\dag} \tau_2 T = - \tau_3,
    T^{\dag} \tau_3 T = \tau_1.
   \]
   After using this $T$ to change the basis for the last
   three spinor indices for fermionic coordinates, the gamma
   matrices become
   \bea
   \label{GMM}
    \nonumber
\gamma^0 = \unit \otimes \unit \otimes \unit
   \otimes \unit &,&
\gamma^1 = -\tau_2 \otimes \tau_3 \otimes \tau_3
   \otimes \tau_3, \\ \nonumber
\gamma^2 = -\tau_1 \otimes \unit \otimes \tau_3
   \otimes \tau_3 &,&
\gamma^3 = -\tau_3 \otimes \unit \otimes \tau_3
   \otimes \tau_3, \\ \nonumber
\gamma^4 = -\tau_2 \otimes \tau_2 \otimes \unit
    \otimes \tau_3 &,&
\gamma^5 = \tau_2 \otimes \tau_1 \otimes \unit
    \otimes \tau_3, \\ \nonumber
\gamma^6 = \unit \otimes \tau_3 \otimes \tau_2
     \otimes \tau_3 &,&
\gamma^7 = -\unit \otimes \tau_3 \otimes \tau_1
      \otimes \tau_3, \\
\gamma^8 = -\unit \otimes \unit \otimes \unit
      \otimes \tau_2 &,&
\gamma^9 = \unit \otimes \unit \otimes \unit
      \otimes \tau_1.
   \eea
   The sixteen-component fermionic coordinate $\lambda$ is
   thus complexified, as an eight-component complex spinor.
   The orbifolding conditions for fermionic variables read
   \bea
   \hat{U}_1^{\dag} \lambda \hat{U}_1
   = \omega^{\sigma_{45} - \sigma_{67}} \lambda
    &,&
   \hat{U}_2^{\dag} \lambda \hat{U}_2
   = \omega^{\sigma_{45} - \sigma_{89}} \lambda,
   \eea
   where $\sigma_{IJ}=i[\gamma^J,\gamma^I]/4$ are the rotational
   generators for spinors:
   \bea
   \nonumber
   \sigma^{45} = \frac{1}{2} \unit \otimes \tau_3 \otimes
   \unit \otimes \unit, \\
   \nonumber
   \sigma^{67} = \frac{1}{2} \unit \otimes \unit \otimes
   \tau_3 \otimes \unit, \\
   \sigma^{89} = \frac{1}{2} \unit \otimes \unit \otimes
   \unit \otimes \tau_3.
   \eea

  Quantum numbers of each variable are summarized in
   Table~\ref{tab}. In this table, $J_{IJ}=M_{IJ}$ for
   vectors, $J_{IJ}=\sigma_{IJ}$ for spinors, and
   remember that $(\lambda^{s_0,t_1,t_2,t_3})^\dag
   =\lambda^{s_0,-t_1,-t_2,-t_3}$.

   \begin{table}[h]
    \begin{center}
    \caption{Quantum Numbers of Variables in
     the BFSS Matrix Theory upon Orbifolding}
    \label{tab}
    \begin{tabular}{|l||c|c|c|c|c|}\hline\hline
     & $J_{45}$ & $J_{67}$ & $J_{89}$ &
     $J_{45}-J_{67}$ & $J_{45}-J_{89}$\\
     \hline\hline
     $Y^{0,1,2,3}$ & 0 & 0 & 0 & 0 & 0 \\ \hline
     $Z^1$ & -1 & 0 & 0 & -1 & -1 \\ \hline
     $Z^2$ & 0 & -1 & 0 & 1 & 0 \\ \hline
     $Z^3$ & 0 & 0 & -1 & 0 & 1 \\ \hline\hline
     $\lambda^{s_0+++}$ & 1/2 & 1/2 & 1/2 & 0 & 0 \\ \hline
     $\lambda^{s_0++-}$ & 1/2 & 1/2 & -1/2 & 0 & 1 \\ \hline
     $\lambda^{s_0+-+}$ & 1/2 & -1/2 & 1/2 & 1 & 0 \\ \hline
     $\lambda^{s_0-++}$ & -1/2 & 1/2 & 1/2 & -1 & -1 \\ \hline\hline
    \end{tabular}
    \end{center}
   \end{table}

\begin{figure}[hbtp]
    \centering
    \psfrag{l1}[][]{$\lambda^{-++}$}
    \psfrag{l2}[][]{$\lambda^{+-+}$}
    \psfrag{l3}[][]{$\lambda^{++-}$}
    \psfrag{l0}[][]{$\lambda^{+++}$}
    \psfrag{Z1}[][]{$Z^1$}
    \psfrag{Z2}[][]{$Z^2$}
    \psfrag{Z3}[][]{$Z^3$}
    \psfrag{Z0}[][]{$Y_{0,1,2,3}$}
    \includegraphics[width=0.60\textwidth]{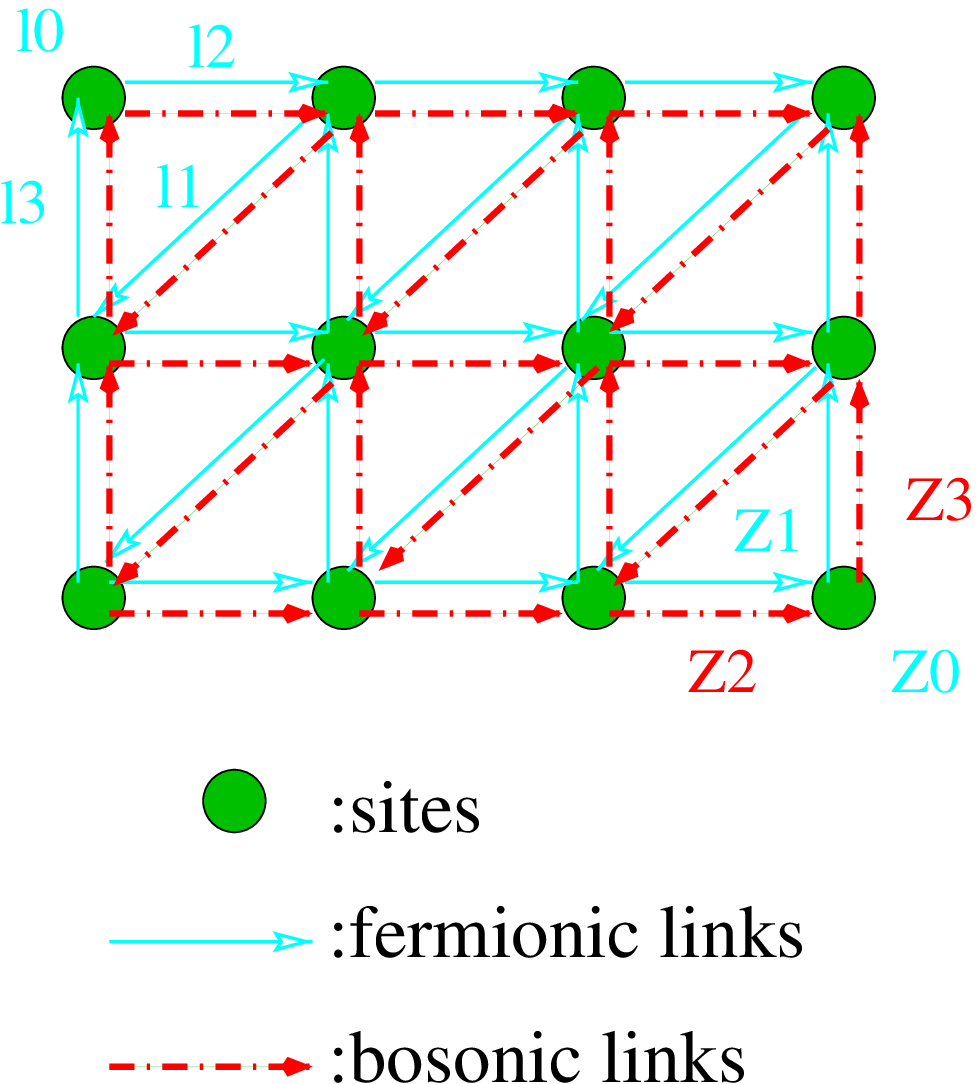}
\caption{Quiver diagram for the Matrix Theory on $\complex^3/\integer_N^2$}
    \label{QDG}
   \end{figure}

   Schematically, the field contents of our orbifolded M(atrix)
   Theory is encoded by a quiver diagram in the form of a
   planar triangular lattice. Figure \ref{QDG} illustrates
   six cells in the quiver diagram. Each variable is a
   $KN^2$-by-$KN^2$ matrix, which can also be viewed as a
   $K$-by-$K$ matrix field living on the sites or links in
   the quiver diagram with $N^2$ sites. The matrix block form
   for the $N^2$-by-$N^2$ indices of these variables is
   dictated by the quiver diagram. From either Table~\ref{tab}
   or Figure \ref{QDG}, we can read off that $Y^{0,1,2,3}$
   and $\lambda^{s_0+++}$, hence $\lambda^{s_0---}$, are
   defined on the sites, while $Z^{1,2,3}$ and other
   fermionic coordinates, as well as their hermitian
   conjugate, on the links \cite{Kaplan}. In the following,
   we will often suppress the $K$-by-$K$ indices, while using
   a pair of integers $(m,n)$, $m,n=1,2,\cdots, N$, to label
   the lattice sites in the quiver diagram. Note that for
   convenience the ``background'' in the diagram is drawn
   like a regular lattice, but remember that at this moment
   no notion of metric has been defined.

   The orbifolded action, as a quiver mechanical model, reads
   \bea
   \label{BFSS1}
    \nonumber
    S &=& \int d\tau Tr \{
{1\over 2R_{11}}[D_\tau, Y^i]^2
 + {1\over 2R_{11}}[D_\tau, Y^m]^2 \\
    \nonumber
&& + \frac{R_{11}}{4}[Y^i,Y^j]^2 + \frac{R_{11}}{2} [Y^i,Y^m]^2 +
     \frac{R_{11}}{4}[Y^m,Y^n]^2 \\
&& - {i\over 2}\lambda^{\dag}[D_\tau,\lambda] + \frac{R_{11}}{2} \lambda^{\dag} \gamma^i [Y_i,\lambda]
     + \frac{R_{11}}{2} \lambda^{\dag} \gamma^m [Y_m,\lambda]
     \}
   \eea
   where $i,j$ run from $1$ to $3$, $m,n$ from $4$ to $9$. Or
   equivalently in terms of the complex coordinates $Z^a$,
   \bea
    \nonumber
    S &=& \int d\tau Tr \{
{1\over 2R_{11}}[D_\tau, Y^i]^2
+ \frac{R_{11}}{4}[Y^i,Y^j]^2 \\
    \nonumber
     && + {1\over R_{11}}[D_\tau, Z^a][D_\tau, Z^{a\dag}]
+ \frac{R_{11}}{2} ([Z^a,Z^{a^\prime \dag}][Z^{a\dag},Z^{a^\prime}]
     + [Z^a,Z^{a^\prime}][Z^{a\dag},Z^{a^\prime \dag}]) \\
    \nonumber
     &&  + R_{11}[Y^i,Z^a][Y^i,Z^{a\dag}] \\
&& - {i\over 2}\lambda^{\dag}[D_\tau,\lambda] + \frac{R_{11}}{2} \lambda^{\dag} \gamma^i [Y_i,\lambda] +
\frac{R_{11}}{\sqrt{2}}
  \lambda^{\dag} (\tld{\gamma}_a [Z^a,\lambda]
+ \tld{\gamma}_a^{\dag} [Z^{a\dag},\lambda] )
     \} \, .
   \label{OA}
   \eea
   In this action, the trace $Tr$ is taken on a
   $KN^2$-by-$KN^2$ matrix, and  $\tld{\gamma}_a =
   (\gamma^{2a+2} - i\gamma^{2a+3})/2$, with
   \bea\label{GMMC}
    \nonumber
\tld{\gamma}^1 &=& -\epsilon \otimes \tau_- \otimes \unit \otimes \tau_3,
\\ \nonumber \tld{\gamma}^2 &=& i\unit \otimes \tau_3 \otimes
\tau_-   \otimes \tau_3, \\
\tld{\gamma}^3 &=& -i \unit \otimes \unit \otimes \unit \otimes \tau_-,
   \eea
   and $\tau_- = (\tau_1 - i \tau_2)/2$.

   Recall that the number of surviving supersymmetries
   corresponds to the number of fermionic variables on each
   site; so we have four supercharges corresponding to
   $\lambda^{s_0+++}$ and $\lambda^{s_0---}$ with $(s_0=1,2)$.
   Relabel fermionic variables according their statuses in
   the quiver
   diagram, $\Lambda_0^s = \lambda^{s+++}$,
   $\Lambda_1^s = \lambda^{s-++}$,
   $\Lambda_2^s = \lambda^{s+-+}$,
   $\Lambda_3^s = \lambda^{s++-}$. Hence, the Yukawa
   interactions in the action~(\ref{OA}) can be recast
   into a tri-linear form:
   \bea
   \nonumber
    S_{BF}&=& \int d\tau \sqrt{2}R_{11} Tr\{
-\Lambda^{\dag}_1\epsilon[Z^1, \Lambda_0] + i \Lambda_2^{\dag} [Z^2,\Lambda_0]
- i \Lambda_3^{\dag} [Z^3,\Lambda_0] \\
    &&
-   [Z^1, \Lambda_2]\epsilon \Lambda_3 - i \Lambda_1 [Z^2,\Lambda_3]
     + i \Lambda_1 [\Lambda_2, Z^3] + h.c.
    \}.
   \label{F}
   \eea
   As well-known in the literature on deconstruction, the
   allowed Yukawa interactions are such that a closed
   triangle (which may a degenerate one) is always
   formed by the two fermionic legs and one bosonic leg
   or site, due to the orbifolded gauge invariance.

  \section{Constructing $d=2+1$ World-volume Theory}
  \label{SEC4}

   Previously in M(atrix) Theory on a 2-torus, the
   compactified coordinates that solve the quotient
   conditions were expressed as covariant derivatives,
   resulting in a $(2+1)$-dimensional SYM on the dual
   torus with 16 supersymmetries. In this section,
   to provide a (de)construction of this $d=2+1$ SYM,
   in the same spirit as advocated in Ref.~\cite{DW}
   and in Sec.~\ref{SEC2}, we will assign non-vanishing
   VEV to the bosonic link variables in our quiver
   mechanical model with the action (\ref{OA}),
   and show that it will approach in the continuum
   limit (i.e., a large $N$ limit) to the desired SYM.
   Though the outcome may be expected in advance (see
   for example \cite{Kaplan}), to see how the limit is
   achieved with a generic triangular lattice and how
   it could be reduced to a rectangular lattice should
   be helpful for our later discussion for wrapping
   matrix membranes.

   \subsection{Moduli and Parametrization of Fluctuations}

   To proceed, we first note that the orbifolding
   conditions (\ref{OCB}) require the bi-fundamental
   bosonic matrix variables be of a particular
   block form in the site indices:
   $(Z^a)_{mn,m'n'} = z^a(m,n) (\hat{V}_a)_{mn,m'n'}$,
   with
   \bea
   (\hat{V}_2)_{mn,m'n'}&=&(V_N)_{m,m'} (\unit_N)_{n,n'},
       \nonumber \\
   (\hat{V}_3)_{mn,m'n'}&=&(\unit_N)_{m,m'} (V_N)_{n,n'},
   \eea
   and $\hat{V}_1:=\hat{V}_2^\dag \hat{V}_3^\dag$. Each
   $z^a(m,n)$ $(a=1,2,3)$ (for fixed $(m,n)$) is a
   $K$-by-$K$ matrix. (Since $\hat{V}_2, \hat{V}_3$ can
   be viewed as the generators of the discrete group
   $\integer_N \otimes \integer_N$, the orbifolded matrices
   are of the form of the so-called {\em crossed product}
   of $Mat(K)$ with $\integer_N^2$ \cite{Connes}\cite{DG},
   expressing the
   properly projected form of the matrices upon orbifolding.)

   In our approach, the key step after orbifolding
   is to assign nonzero VEV to each element
   $z^a(m,n)$:
   \[
    z^a = \bra z^a \ket + \tld{z}^a,
   \]
   with $\bra z^a \ket = c_a/\sqrt{2}$, $c_a$ complex
   numbers to be specified later. Later in this paper
   we will take the following moduli:
   \be\label{SETTING}
    c_1=0,c_2=NR_2/2\pi, c_3=NR_3/2\pi.
   \ee
   Accordingly, the fluctuations can be parameterized as
   \bea  \label{COO1}
   \tld{z}^1 &=& (\phi_1 + i\phi_1^\prime)/\sqrt{2},
   \nonumber \\
   \tld{z}^2 &=& (\phi_2 - iR_2A_2)/\sqrt{2},
   \nonumber \\
   \tld{z}^3 &=& (\phi_3 - iR_3A_3)/\sqrt{2}.
   \eea
   We remark that all the new variables appearing in the
   parametrization are $K$-by-$K$ matrices and that the
   parametrization (\ref{COO1}) is intimately related with
   our choice (\ref{SETTING}) for the moduli.

   The above decomposition has an obvious physical
   interpretation: $\bra z^a\ket$ are the modulus
   part and $\tld{z}^a$ the fluctuations. For the
   modulus part, as in Eq.~(\ref{WS}) for our previous
   deconstruction of IIA string, we expect that the
   lattice constants are simply inversely proportional
   to the VEV of the bi-fundamental boson variable.
   Schematically $a \propto 1/c_{2,3}$ such that $a N\sim 1$.
   Indeed we will see that the spatial geometry, after
   taking the continuum limit, emerges from three modulus
   parameters $\bra z^a \ket$ ($a=1,2,3$); in other words,
   a world-volume metric will be constructed out of them.
   Our choice (\ref{SETTING}) of moduli will greatly
   simplify the complexity that arises from our quiver
   diagram being a two dimensional triangular lattice.
   For the fluctuations, the parametrization in Eq.
   (\ref{COO1}) corresponds to the infinitesimal form
   of the polar-coordinate decomposition of $z^a$. Note
   that the Yang-Mills fields come from the imaginary
   (or angular) part of the bi-fundamental fluctuations.

   Before proceeding to consider the continuum limit, we
   make the following remarks on algebraic properties of
   the matrices $\hat{V}_a$. Let $f$ be a diagonal matrix
   in the site indices:
   $f_{mn,m'n'} = f(m,n)\delta_{mm'}\delta_{nn'}$,
   then
   \be
   \hat{V}_a^{\dag} f \hat{V}_a = S_a f,
   \ee
   where $S_a$ is the shift operator by a unit
   along $a$-direction, i.e.
   \bea
    S_1 f(m,n) &=& f(m-1,n-1), \\
    S_2 f(m,n) &=& f(m+1,n), \\
    S_3 f(m,n) &=& f(m,n+1).
   \eea
   Subsequently, one can easily derive the following
   relations for manipulations on a site function $f$:
   \be
   \hat{V}_a f \hat{V}_a^{\dag} = S_a^{-1}f,\quad
   \hat{V}_a (S_af - f) = [f, \hat{V}_a], \quad
   [f, \hat{V}_a^{\dag}] = -[f,\hat{V}_a]^{\dag}.
   \ee

\subsection{World-volume Geometry from Deconstruction}

 In the following two subsections, we will examine
 the bosonic part of the action,
   \bea
    \nonumber
    S_B &=& \int d\tau Tr \{
    {1\over R_{11}}|[D_\tau, Z^a]|^2 -
    \frac{R_{11}}{2}(|[Z^a,Z^{a^\prime
    \dag}]|^2 + |[Z^a,Z^{a^\prime}]|^2) \\
&& + {1\over 2R_{11}}[D_\tau, Y^i]^2 - R_{11}|[Y^i,Z^a]|^2 +
    \frac{R_{11}}{4}[Y^i,Y^j]^2\}.
   \eea
   With the parametrization proposed above, the
   quadratic part for $Y^i$ becomes
   \bea
    \nonumber
    S_Y &=& \int d\tau Tr \{
    {1\over 2R_{11}}[D_\tau, Y^i]^2 - R_{11}|[Y^i,Z^a]|^2\} \\
    &=& \int d\tau Tr\{{1\over 2R_{11}}
    |\dot{Y}^i + i[Y_0, Y^i]|^2
    -R_{11}\big| \partial_a Y^i \frac{2\pi S_a(z^a)}{N}
    + [Y^i,S_a(\tld{z}^a)]\big|^2\}
   \eea
   where we have introduced the lattice partial derivative,
   \be
   \partial_a f := N (S_a f- f)/2\pi.
   \ee

   First we see the information of the metric on the
   would-be world-volume is contained in the term
   \be
   \label{KE}
    S_{YK} = \int d\tau R_{11} Tr \{
    {1\over 2R_{11}^2}[\dot{Y}^i]^2 -
    \frac{(2\pi)^2|S_a(z^a)|^2}{N^2}[\partial_a Y^i]^2\}.
   \ee
   Recall that the target space index $i$ runs from 1
   to 3 and that the index $a$ also from 1 to 3. Note
   that $\partial_1$ is not independent of
   $\partial_{2,3}$, because of the relation
   \[
    \partial_1 = -S_2^{-1}S_3^{-1}\partial_2 - S_3^{-1}\partial_3.
   \]
   Changing the unit for time $\tau':=R_{11} \tau$ and then
   suppressing the prime for $\tau'$, Eq.~(\ref{KE}) changes
   into
   \be
    S_{YK} = \int d\tau {1\over 2} Tr \{
    g^{00}[\dot{Y}^i]^2 - g^{22}[\partial_2 Y^i]^2
    -g^{33}[S_2\partial_3 Y^i]^2
    -g^{23}\{\partial_2 Y^i, S_2\partial_3 Y^i\}\},
   \ee
   from which we can read off the contravariant metric
   on world-volume
   \bea \nonumber\label{WVM}
    g^{00}=1,
    g^{22} = \frac{8\pi^2(|z^1|^2+|S_2(z^2)|^2)}{N^2},\\
    g^{33} = \frac{8\pi^2(|z^1|^2+|S_1^{-1}(z^3)|^2)}{N^2},
    g^{23} = \frac{8\pi^2|z^1|^2}{N^2}
   \eea
   and $g^{0 k}=0$ for the spatial index $k=2,3$. In this
   way, we see how the world-volume naturally metric
   arises from our (de)construction.

   Now we are ready to take the continuum limit. First regularize
   the trace $Tr$ to be $\sum (2\pi)^2 tr/N^2\kappa$; with our
   choice (\ref{SETTING}) of the moduli, we take $\kappa = R_2R_3$.
   To see how world-volume geometry comes out of this construction,
   we consider the continuum limit with $N\to \infty$, then
   $Tr \rightarrow \int d^2\sigma tr/R_2R_3$, where $tr$ is the
   trace over $K$-by-$K$ matrices. Here we have taken the spatial
   world-volume coordinates $\sigma^{2,3}$ to run from $0$ to $2\pi$.
   Subsequently, the action in the continuum limit reads
   \be
   \label{YK}
   S_{YK} = - \int d\tau \int d^2\sigma {1\over R_2R_3}
   tr \{ {1\over 2} g^{\alpha\beta}
   \partial_\alpha Y^{i} \partial_\beta Y^{i} \},
   \ee
   in which the contravariant metric on $(2+1)$-dimensional
   world-volume is
   \be\label{WVMC}
    (g^{\alpha\beta})=diag\; (-1, R_2^2, R_3^2).
   \ee
   In fact, we have specified the behavior of fluctuations
   $\tld{z}^a$, in the continuum limit $N\rightarrow \infty$,
   as ${\mathcal O}(1)$; namely the contribution from the
   fluctuations to the world-volume metric in Eq.~(\ref{WVM})
   smears or smoothes in the large-$N$ limit. As another fact,
   due to our choice $c_1=0$, world-volume metric (\ref{WVM})
   becomes diagonal, as shown in Eq. (\ref{WVMC}). Accordingly,
   the factor $\kappa$, taken to be $R_2R_3$, multiplied by
   the coordinate measure $d^2\sigma$, is nothing but the
   invariant world volume measure, namely
   $\kappa = \sqrt{-det(g_{\alpha\beta})}$, where
   \be
   \label{CVMETRIC}
    (g_{\alpha\beta})=diag\; (-1, 1/R_2^2, 1/R_3^2)
   \ee
   A further check shows that the area of the
   world-volume torus is
   \bea\label{AREA}
   &&\mbox{\em Area of world-volume torus}
   \nonumber \\
   &=& \sqrt{-det(g_{\alpha\beta})}\cdot
   \mbox{\em coordinate area} =\frac {(2\pi)^2}{R_2 R_3}.
   \eea

   Now for the continuum limit of the full $S_{Y}$, it is
   easy to show that
   \be
   \label{YGK1}
    S_Y = \int d\tau d^2\sigma {1\over 2R_2R_3} tr \{
   -g^{\alpha\beta}D_\alpha Y^iD_\beta Y^i
   + [\phi_1^\prime, Y^i]^2 + [\phi_a, Y^i]^2\}
   \ee
   where $D_\alpha = \partial_\alpha + i[A_\alpha,.]$ and $A_0=Y^0$.

   In this subsection we have constructed, from our quiver
   mechanics, the toroidal geometry on the spatial world-volume,
   as well as the standard gauge interactions of world-volume
   scalar fields. (Incidentally let us observe that besides
   the local world-volume geometry, the modular parameter that
   describes the shape of the compactified 2-torus in target
   space is also dictated by the moduli $\bra z^a\ket$ that also fix
   the shape of the triangular unit cell in the quiver diagram.
   However, in this paper we will concentrate on our special
   choice ~(\ref{SETTING}) for the moduli, and leave the
   analysis of more general moduli to the sequential paper.)

   \subsection{Yang-Mills Field Construction}

   Now we consider the action of bi-fundamental fields; our
   goal here is to construct the desired SYM.

   Collecting the relevant terms in the action, we have
   \be
   \label{SZ}
    S_Z = \int d\tau Tr \{
{1\over R_{11}}|[D_\tau, Z^a]|^2 - \frac{R_{11}}{2}(|[Z^a,Z^{a^\prime
    \dag}]|^2 + |[Z^a,Z^{a^\prime}]|^2)\}.
   \ee
   Write $[Z^{a^\prime\dag}, Z^a] =
   \hat{V}_{a^\prime}^\dag P_{a^\prime a} \hat{V}_a$,
   with $P_{a^\prime a} = z^{a^\prime\dag} z^a
   - S_{a^\prime}^{-1}z^a S_a^{-1}z^{a^\prime\dag}$.
   Separating the VEV and fluctuations,
   \[
   P_{a^\prime a} = S_{a^\prime}^{-1}\partial_{a^\prime}\tld{z}^a
   \frac{2^{-1/2}\bar{c}_{a^\prime} +
   S_a^{1}\tld{z}^{a^\prime \dag}}{(2\pi)^{-1}N}
   +\frac{2^{-1/2}c_a+\tld{z}^a}{(2\pi)^{-1}N}S_a^{-1}
   \partial_a\tld{z}^{a^\prime\dag}
   +[\tld{z}^{a^\prime\dag}, \tld{z}^a].
   \]
   Therefore, $P_{a^\prime a}^\dag = P_{aa^\prime}$.
   Moreover, $[Z^a,Z^{a^\prime}] = Q_{a^\prime a}
   \hat{V}_a\hat{V}_{a^\prime}$, with
   $Q_{a^\prime a} = z^aS_a^{-1}z^{a^\prime}-
   z^{a^\prime}S_{a^\prime}^{-1}z^a$; thus
   \[
Q_{a^\prime a} = S^{-1}_{a^\prime}\partial_{a^\prime} \tld{z}^a \frac{2^{-1/2}c_{a^\prime} +
S_a^{-1}\tld{z}^{a^\prime}}{(2\pi)^{-1}N} -\frac{2^{-1/2}c_a +S_{a^\prime}^{-1}\tld{z}^a} {(2\pi)^{-1}N}S_a^{-1}
\partial_a\tld{z}^{a^\prime}
    +[S_{a^\prime}^{-1}\tld{z}^a, \tld{z}^{a^\prime}].
   \]
   Hence $Q_{a^\prime a} = - Q_{aa^\prime}$. The action ~(\ref{SZ})
   is recast into
   \bea
   \nonumber
    S_Z &=& \int d\tau Tr \{
    {1\over R_{11}}|[D_\tau, Z^a]|^2 - \frac{R_{11}}{2}
    (\sum \limits_{a} P_{aa}^2 + 2\sum\limits_{a^\prime
    <a}(|P_{a^\prime a}|^2
    + |Q_{a^\prime a}|^2)) \}\\
&= & \int d\tau Tr \{|S_{a0}|^2 - ({1\over 2}\sum \limits_{a} P_{aa}^2 + \sum\limits_{a^\prime <a} (|P_{a^\prime
a}|^2 + |Q_{a^\prime a}|^2))\}
   \eea
   where, in the second line, a rescaling of time is implied,
   and $S_{a0} = 2\pi S_a^{-1}\partial_a A_0 z^a/N
   - i \dot{\tld{z}}^a + [S_a^{-1} A_0,\tld{z}^a]$.
   Now we assign the VEV (\ref{SETTING}), and calculate the
   following intermediate quantities:
   \bea
    S_{10} &=& (-iD_0\phi_1 - D_0\phi_1^\prime)/\sqrt{2},\\
    S_{k0} &=& (R_kF_{k0} - iD_0\phi_k)/\sqrt{2},\\
    P_{11} &=& i[\phi_1,\phi_1^\prime],\\
    P_{kk} &=& R_kD_k\phi_k,\\
P_{1k} &=& (R_kD_k\phi_1 - i R_kD_k\phi_1^\prime
   + [\phi_1,\phi_k] - i [\phi_1^\prime, \phi_k])/2 ,\\
Q_{1k} &=& (-R_kD_k\phi_1 - i R_kD_k\phi_1^\prime
   - [\phi_1,\phi_k] - i [\phi_1^\prime, \phi_k])/2 ,\\
P_{23} &=& (R_2D_2\phi_3 + R_3D_3\phi_2
   + [\phi_2,\phi_3] - iR_2R_3 F_{23})/2 ,\\
Q_{23} &=& (R_2D_2\phi_3 - R_3D_3\phi_2
   - [\phi_2,\phi_3] - iR_2R_3 F_{23})/2
   \eea
   and their squares
   \bea
    |S_{10}|^2 &=& (|D_0\phi_1|^2 + |D_0\phi_1^\prime |^2)/2,\\
    |S_{k0}|^2 &=& (R_k^2|F_{k0}|^2 + |D_0\phi_k|^2)/2,\\
    |P_{1k}|^2 &=& (|R_kD_k\phi_1 - i [\phi_1^\prime, \phi_k]|^2
+ |R_kD_k\phi_1^\prime + i[\phi_1,\phi_k]|^2)/4 ,\\
    |Q_{1k}|^2 &=& (|R_kD_k\phi_1 + i [\phi_1^\prime, \phi_k]|^2
+ |R_kD_k\phi_1^\prime - i[\phi_1,\phi_k]|^2)/4 ,\\
    |P_{23}|^2 &=& (|R_2D_2\phi_3 + R_3D_3\phi_2|^2
+ |[\phi_2,\phi_3] - iR_2R_3 F_{23}|^2)/4 ,\\
    |Q_{23}|^2 &=& (|R_2D_2\phi_3 - R_3D_3\phi_2|^2
+ |[\phi_2,\phi_3] + iR_2R_3 F_{23}|^2)/4.
   \eea
   Here the gauge field strength $F_{\alpha\beta}$ is defined
   in the usual way. Subsequently, we have, in the continuum limit,
   \be\label{YMS}
    S_Z = \int d\tau d^2\sigma {1\over R_2R_3} tr\{
    -{1\over 4} F^{\alpha\beta}F_{\alpha\beta}
    -{1\over 2} g^{\alpha\beta}D_\alpha \phi^M D_\beta \phi^M
    + {1\over 4} [\phi^M,\phi^N]^2\},
   \ee
   in which $\phi^M$ include $\phi_{1,2,3}$ and $\phi_1^\prime$.

   In summary, Eqs.~(\ref{YGK1}) and (\ref{YMS}) leads to the
   complete continuum bosonic action:
    \be
    S_B = \int d\tau d^2\sigma {1\over R_2R_3} tr\{
    -{1\over 4} F^{\alpha\beta}F_{\alpha\beta}
    -{1\over 2} g^{\alpha\beta}D_\alpha \Phi^I D_\beta \Phi^I
    + {1\over 4} [\Phi^I,\Phi^I]^2\}
   \label{BYM3}
   \ee
   where $\Phi^I$ include both $\phi^M$ and $Y^i$.

   Eq.~(\ref{BYM3}) is the dimensional reduction from the
   bosonic part of either four-dimensional $\natural=4$
   SYM or ten-dimensional $\natural=1$ SYM. Therefore,
   with appropriate rescaling of world-volume fields,
   the action ~(\ref{BYM3}) possesses an $O(7)$ symmetry
   for fields $Y^i$, $\phi_a$ and $\phi_1^\prime$.

   We remark that the first two terms in the continuum action
   (\ref{BYM3}) contain one and the same world-volume metric.
   This can not be a mere coincidence; it is dictated by the
   stringent internal consistency of our deconstruction
   procedure.

   \subsection{Fermion Construction}

   The deduction of the continuum action for fermionic
   variables can be carried out in a similar way. We
   only emphasize that the phase of D-particle moduli
   can be absorbed int redefinition of fermionic variables
   in Eq.~(\ref{F}). Indeed after assigning non-zero VEV
   (or moving the D-particles collectively away from the
   orbifold singularity), the free fermionic action becomes
   \bea
   \nonumber
   S_{FF} &=& \int d\tau \frac{(2\pi)^2}{N^2\kappa}
   \sum\limits_{m,n} tr \{
   \lambda^{1\dag}_{m,n} (-)\epsilon c_1
   (\lambda^0_{m+1,n+1}-\lambda^0_{m,n}) \\
   \nonumber
   && + \lambda^{2\dag}_{m,n} i c_2 (\lambda^0_{m-1,n}
   -\lambda^0_{m,n}) + \lambda^{3\dag}_{m,n}
   (-i) c_3 (\lambda^0_{m,n-1}-\lambda^0_{m,n}) \\
   \nonumber
   && - (\lambda^2_{m+1,n}-\lambda^2_{m,n-1}) \epsilon c_1
     \lambda^3_{m,n} + i \lambda^1_{m,n}
    c_2 (\lambda^3_{m+1,n+1}-\lambda^3_{m,n+1})\\
   && + i \lambda^1_{m,n} c_3
     (\lambda^3_{m+1,n+1}-\lambda^3_{m+1,n}) + h.c.
    \}
   \label{FERM}
   \eea
   where we have introduced $\lambda_a$ by writing
   $(\Lambda_a)_{mn,m'n'} = (\lambda_a)_{mn}
   (\hat{V}_a)_{mn,m'n'}$ for $a=0,1,2,3$, with
   $\hat{V}_0$ the unit matrix. With the choice
   (\ref{SETTING}) and $\kappa=R_2R_3$, the continuum
   limit reads
   \bea
   \nonumber
    S_{FF} &=& \int d\tau d^2\sigma {1 \over R_2R_3}
    tr\{\lambda^{\dag}_2 (-i) R_2\partial_2 \lambda_0
    +\lambda^{\dag}_3 i R_3\partial_3 \lambda_0
    \\
    && \lambda_1 iR_2\partial_2 \lambda_3
    + \lambda_1 iR_3\partial_3 \lambda_2 + h.c.
    \}.
   \label{FMN}
   \eea
   Comparing with our definition of gamma matrices,
   (\ref{GMM}) and (\ref{GMMC}), we get
   \be
   S_{FF} = \int d\tau d^2\sigma {1 \over R_2R_3}
   tr\{ \lambda^\dag  i(\gamma^7R_2\partial_2
     +\gamma^9R_3\partial_3)\lambda\}/2.
   \ee
   The appearance of the seventh and ninth gamma
   matrices can be easily understood from the
   parametrization of the fluctuations in
   Eq.~(\ref{COO1}), since the gauge fields arise
   from the fluctuations in the imaginary part of
   complex coordinates $\tld{z}^a$.

   The continuum limit of the complete action for
   fermionic variables is thus
   \bea
   \label{COMFER}
   \nonumber
   S_F&=&\int d\tau d^2\sigma {1\over R_2R_3}tr\{
   -{i\over 2} \lambda^\dag[D_\tau,\lambda]
   +{i\over 2}\lambda^\dag\gamma^7R_2[D_2,\lambda]
   +{i\over 2}\lambda^\dag\gamma^9R_3[D_3,\lambda]\\
   && +{1\over 2}\lambda^\dag\gamma^I[\Phi^I,\lambda]
   \}.
   \eea

   Summarily, Eq.~(\ref{BYM3}) plus Eq.~(\ref{COMFER})
   constitute precisely the desired $d=2+1$ SYM with
   sixteen supercharges. This outcome in the continuum
   limit verifies our idea that M(atrix) Theory
   compactification can be deconstructed in terms
   of orbifolding and quiver mechanics, so that we
   are on the right track to IIB/M duality. Moreover,
   the constructed world-volume geometry, in view of
   world-volume/target-space duality, provides the
   platform for the discussion in next section on
   the matrix membranes wrapping on the compactified
   2-torus in target space.

  \section{Wrapping Membranes and Their Spectrum}
  \label{SEC5}

   In the above we have ``(de)constructed" the M(atrix)
   Theory compactified on a 2-torus, by starting from
   orbifolding and then assigning non-zero VEVs to
   bi-fundamental bosons. According to the IIB/M duality
   conjecture, M(atrix) Theory compactified on a
   2-torus is dual to IIB string theory on a circle.
   In this section, we will show that the quiver
   mechanics resulting from our deconstruction procedure
   indeed possesses features that are characteristic to
   the IIB/M duality. More concretely, we will explicitly
   construct matrix states in our quiver mechanics, which
   in the continuum limit correspond to membranes wrapping
   over compactified a 2-torus in the target space. And we
   will show a $SL(2,\integer)$ symmetry for these states
   and the emergence from these states of a new flat
   dimension when the compactified 2-torus shrinks to
   zero. As mentioned in Sec.~\ref{SEC1}, these features
   are central for verifying the IIB/M duality conjecture.
   The existence of these states and their properties were
   discussed in the literature of M(atrix) Theory in the
   context of $d=2+1$ SYM, but to our knowledge the
   explicit matrix construction of these states is still
   absent in the literature.

   \subsection{Classical Wrapping Membrane}

   To motivate our construction, we recall that a
   closed string winding around a circle is described by
   a continuous map from the worldsheet circle to the
   target space circle, satisfying the periodic condition:
   $x(\sigma + 2\pi) = x(\sigma) + 2\pi w R$, where $w$
   is the winding number. For a $\complex^2/\integer_N$
   orbifold, a winding string is a state in the twisted
   sector satisfying $z(\sigma + 2\pi) = e^{i2\pi w/N}
   z(\sigma)$.

   In the same way, in the non-linear sigma model
   for a toroidal membrane wrapping on a compactified
   2-torus, we have the boundary conditions
   \bea
    \nonumber
    x_1(p+2\pi,q)=x_1(p,q)+2\pi m R_1 &,&
    x_1(p,q+2\pi)=x_1(p,q)+2\pi n R_1,\\
    x_2(p+2\pi,q)=x_2(p,q)+2\pi s R_2 &,&
    x_2(p,q+2\pi)=x_2(p,q)+2\pi t R_2
   \eea
   where $(p,q)$ ($0\le p,q \le2\pi$) parameterize
   the membrane. The solutions to these conditions
   are given in Eq.~(\ref{WDM}). They contain four
   integers $(m,n,s,t)$. One combination of them is
   $SL(2,\integer)$ invariant, i.e. the {\em
   winding/wrapping number}:
   \be
    w = mt-ns.
   \ee
   (This is a slight modification of Schwarz's original
   construction \cite{SCH}\cite{SCHRE}; note that this combination
   appeared also in the context of noncommutative torus,
   see for example Eq.~(4.2) in \cite{REVNCG}). The
   geometric significance of $w$ is simply the ratio of
   the pullback area of the membrane to the area of the target torus. In IIB/M
   duality, $w$ is related to the Kaluza-Klein momentum
   in the newly emergent dimension.

   For a membrane wrapping on the orbifold
   $\complex^3/\integer_N^2$,
   the twisted boundary conditions become
   \bea
   \nonumber
    z^2(p+2\pi,q) = e^{i2\pi m/N}z^2(p,q)&,&
    z^2(p,q+2\pi) = e^{i2\pi n/N}z^2(p,q),\\
    z^3(p+2\pi,q) = e^{i2\pi s/N}z^3(p,q)&,&
    z^3(p,q+2\pi) = e^{i2\pi t/N}z^3(p,q),
   \eea
   as well as
   \be
    z^1(p+2\pi,q) = e^{-i2\pi (m+s)/N}z^1(p,q),
    z^1(p,q+2\pi) = e^{i2\pi (n+t)/N}z^2(p,q).
   \ee
   Their solutions are of the form
   \bea\label{SOL}
   \nonumber
    z^1(p,q;\tau)&=& 2^{-1/2} c_1(\tau)
       e^{-i[(m+s)p+(n+t)q]/N}
       \cdot(\mbox{oscillator modes}),\\
   \nonumber
    z^2(p,q;\tau)&=& 2^{-1/2} c_2(\tau) e^{i(mp+nq)/N}
       \cdot(\mbox{oscillator modes}), \\
    z^3(p,q;\tau)&=& 2^{-1/2} c_3(\tau) e^{i(sp+tq)/N}
       \cdot(\mbox{oscillator modes}).
   \eea
   Here the coefficients $c_a (\tau)$ describe the
   center-of-mass degrees of freedom. Again, the
   solutions contain a quadruple, $(m,n,s,t)$, of
   integers, giving rise to an $SL(2,\integer)$
   invariant wrapping number $w = mt-ns$.

   Now let us consider a stretched membrane (\ref{SOL}),
   with the oscillator modes suppressed (by setting
   the corresponding factor to unity). The membrane
   stretching energy density is known
   to be proportional to $\{Y^I,Y^J\}\{Y^I,Y^J\}$,
   in terms of Poisson bracket with respect to the
   canonical symplectic structure on the membrane
   spatial world-volume. (See, e.g., Ref. \cite{Taylor}
   for a latest review and also for a comprehensive
   list of references.) It is easy to show that with
   choosing $c_1=0$, $c_{2,3}\propto NR_{2,3}$, the
   Poisson bracket of $z^2(p,q)$ and $z^3(p,q)$ is
   \be\label{PSBE}
    \{z^2(p,q),z^3(p,q)\} \propto wR_2R_3.
   \ee
   So the energy of this wrapping configuration
   is proportional to $(wR_2R_3)^2$, having the
   correct dependence on $w$ and on the area of
   the 2-torus in target space, because the choice
   of $c_a$ corresponds to a regular target torus.

   We note that to show IIB/M duality in the context
   of M(atrix) Theory, we need a construction of
   wrapping membranes either in the setting of $d=2+1$
   SYM or in that of our quiver mechanics. But the
   above construction (\ref{SOL}) is in neither, so it
   is {\sl not} what we are looking for. However, we
   may view it as a sort of the classical and continuum
   limit of the wrapping matrix membrane we are looking
   for, pointing to the direction we should proceed.

   \subsection{Wrapping Matrix Membranes}

   Now we try to transplant the above continuum
   scenario into the notion of matrix membrane.
   It is well-known that for the finite matrix
   regularization of a membrane, whose topology
   is a torus, one needs to prompt the membrane
   coordinates into the clock and shift matrices,
   namely to substitute $e^{ip}\rightarrow U_K$
   and $e^{iq}\rightarrow V_K$ \cite{Taylor},
   where $K$ is the number of D-particles that
   constitute the membrane. The key property of
   $U_K$ and $V_K$ for this substitution to work
   is
   \be \label{UV}
   U_K^m V_K^n=(\omega_K)^{mn} V_K^n U_K^m,
   \ee
   with arbitrary integers $m,n$, and
   $\omega_K=e^{i2\pi/K}$. In the classical
   or continuum limit, it gives to the correct
   Poisson bracket between $p$ and $q$.
   Eq. (\ref{SOL}) motivates us to deal with
   the fractional powers of the clock and shift matrices.
   The guide line to formulate the arithmetics of the fractional powers
   is to generalize Eq.~(\ref{UV}) to be valid for fractional powers.
   Below we model a simplest substitution scheme
   \be
    e^{ip/N}\rightarrow U_{KN^2}=:U_K^{1/N},
    e^{iq/N}\rightarrow V_{KN^2}=:V_K^{1/N},
   \ee
   which is enough at the current stage
   to enable us to explore the physics of IIB/M duality.
   In accordance with these substitutions,
   we introduce the
   following {\it Ansatz} for the matrix
   configuration of a wrapping membrane:
   \bea
   \label{STT}
    \nonumber
    Z_{(mem)}^1 &=& 2^{-1/2}c_1 U^{-(m+s)} V^{-(n+t)}
         \otimes \hat{V}_1, \\
    \nonumber
    Z_{(mem)}^2&=& 2^{-1/2}c_2 U^{m} V^{n}
         \otimes \hat{V}_2, \\
    Z_{(mem)}^3 &=& 2^{-1/2}c_3 U^{s} V^{t}
         \otimes \hat{V}_3.
   \eea
   with omitted subscripts $U:= U_{KN^2}, V:= V_{KN^2}$.

   These matrices are $KN^4$-by-$KN^4$, of the form of
   a direct product of three factors, with the rank $KN^2$,
   $N$ and $N$ respectively. In our quiver mechanics,
   the first factor describes $K$ D-particles on the
   $\integer_N^2$ orbifold while
   the last two factors describe the deconstructed
   torus coordinates. The form of the last two factors in {\it
   Ansatz} (\ref{STT}), say $\hat{V}_2$ in $Z_{(mem)}^2$,
   is dictated by the orbifolding conditions. The form of
   the first factor, say $U^{m} V^{n}$ in $Z_{(mem)}^2$ is
   motivated by Eqs. (\ref{SOL}). In other words, the
   expression $c_2 U^{m} V^{n}/\sqrt{2}$ can be viewed
   as polar coordinate decomposition of $Z_{(mem)}^2$,
   in which $c_2$ is interpreted as the distance of the membrane on the
   orbifold to the singularity, while the unitary matrix
   $U^{m} V^{n}$ shows how the membrane constituents
   wrap in the angular direction of $Z^2$. A similar
   interpretation holds for $Z_{(mem)}^3$ and $Z_{(mem)}^1$.

   Though the above intuitive picture looks satisfactory,
   the real justification for the {\it Ansatz} (\ref{STT})
   to describe wrapping membranes on a compactified 2-torus
   should come the detailed study of the physical properties
   of these states in accordance with M(atrix) Theory, which
   will be carried out in next a few subsections.

 \subsection{Spectrum and $SL(2,\integer)$ Invariance}

   From now on, we will suppress the subscript in $Z_{(mem)}^a$.
   First let us verify that the equations of motion, that
   follow from the action~(\ref{OA}),
   \be\label{EN}
   \ddot{Z}^a + \frac{R_{11}^2}{2}([Z^b,[Z^{b\dag}, Z^a]]
   +[Z^{b\dag},[Z^b,Z^a]]) = 0.
   \ee
   can be satisfied by the Ansatz (\ref{STT}). Indeed,
   from the experience of the equation of motion
   for the spherical matrix membrane, to solve the
   time-dependence of the scalar
   coefficients ($c_a$ here) is by far kind of
   involved \cite{KT}\cite{Rey_S};
   for the toroidal case, however, this situation is
   simplified because the coefficient functions
   are complex. It is
   a straightforward calculation to show that the equations
   of motion (\ref{EN}) are satisfied if we take
   $c_a(\tau) =c_{a0} e^{-i\omega_a \tau}$,
   with
   \be
   \label{OMEG}
   \omega_a^2 = R_{11}^2(1-\cos {2\pi w\over KN^2})
   \sum\limits_{b\neq a} |c_{b0}|^2.
   \ee
  Consistent with our previous choice ~(\ref{SETTING}) for
  the moduli, hereafter we will just take $c_{10}=0$,
  $c_{k0} = NR_k/2\pi$ for $(k=2,3)$. Accordingly, we have
   \bea \label{OMEG1}
\omega_2^2 &=& ((2\pi)^{-1}NR_{11}R_3)^2
       (1-\cos {2\pi w\over KN^2}), \\
\omega_3^2 &=& ((2\pi)^{-1}NR_{11}R_2)^2 (1-\cos {2\pi w\over KN^2}).
   \eea

   Moreover concerning the membrane dynamics, one can
   easily check an additional constraint equation
   \be
    [\dot{Z}^a,Z^{a\dag}]+[\dot{Z}^{a\dag},Z^a]=0,
   \ee
   which plays the role of the gauge fixing of the membrane
   world-volume diffeomorphism symmetry; see, e.g.,
   ref. \cite{Taylor}.

   Next let us examine the energy of the wrapping
   membrane states (\ref{STT}). By a direct calculation,
   we find that the potential (or stretching) energy
   density ({\sl before} taking the trace $Tr$) is
   proportional to the {\it unit} matrix, $\unit_{N^4}$:
   \be
   \label{ENGD}
   V=R_{11}\sum\limits_{a<b} \{|[Z^a,Z^{b\dag}]|^2+|[Z^a,Z^b]|^2 \}
    =(N/2\pi)^4R_{11}(1-\cos {2\pi w\over KN^2})(R_2R_3)^2\unit_{N^4},
   \ee
   implying a uniform stretching energy density on the membrane.
   We regularize the trace $Tr$ to be $(N^4)^{-1}(2\pi)^2\sum$;
   therefore $Tr\unit_{KN^4}=(2\pi)^2K$.
   Then the total stretching energy is
   \be\label{EW}
    E_w := Tr V = (2\pi)^{-2}KN^4R_{11}(1-\cos {2\pi w\over KN^2})(R_2R_3)^2.
   \ee
   Including the kinetic energy due to the $\tau$-dependence of
   $c_a$, the total energy of the membrane is given by
   \bea\label{ENG}
   \nonumber
    H_{mem} &=&  Tr\{\frac{1}{R_{11}}\sum\limits_{a=1}^3 |\dot{Z}^a|^2 + V \}
    \\\nonumber
    &=& \frac{1}{2R_{11}} \sum_{a=2}^3 |\dot{c}_a|^2 Tr\unit_{N^4}+E_w
    \\
    &=& 2 E_w.
   \eea

   A characteristic feature of Eqs.~(\ref{EW}) and (\ref{ENG})
   is that though the states given by {\it Ansatz} (\ref{STT})
   contain four integers $(m,n;s,t)$, their energy only depends
   on the wrapping number $w=mt-ns$, which is known invariant
   under the $SL(2,\integer)$ transformations
   \be
   \left( \begin{array}{c}
    m'\\n'
   \end{array} \right) =
   \left( \begin{array}{cc}
    a & b\\
    c & d
   \end{array} \right)
   \left(\begin{array}{c}
    m\\n
   \end{array} \right), \qquad
   \left( \begin{array}{c}
    s'\\t'
   \end{array} \right)=
    \left( \begin{array}{cc}
a & b\\
c & d
\end{array} \right)
    \left(\begin{array}{c}
s\\
t
\end{array} \right),
\label{SL2} \ee
   where $a,b,c,d$ are integers, satisfying $ad-bc=1$.
   This result is one chief evidence to justify our
   matrix construction (\ref{STT}), showing the
   $SL(2,\integer)$ symmetry of the spectrum of
   the wrapping membranes directly and explicitly
   in a constructive manner; this $SL(2,\integer)$ in Eqs.~(\ref{SL2}) is right the
   modular symmetry of the target torus, as well being interpreted as the S-duality
   in IIB string theory.

  \subsection{Emergent Flat Dimension}

   In last subsection, we have seen that the stretched
   energy $E_w$ in Eq.~(\ref{EW}) contains a factor
   $(1-\cos (2\pi w/ KN^2))$. Such factor is a common praxis
   in the lattice (gauge) field theory for the spectrum
   of the lattice Laplacian. In the large-$N$ limit,
   when $w$ is small compared to $N^2$, one has
   $(1-cos(2\pi w/KN^2)) \rightarrow 2\pi^2w^2/K^2N^4$.
   Hence $E_w$ recovers the $w^2(R_2R_3)^2$
   behavior of the Poisson bracket calculation
   for continuous membranes in previous subsection 5.1.
  (See Eq.(\ref{PSBE}) and discussion below it).
   More precisely, we have
   \be\label{XX}
   E_w = w^2R_{11}(R_2R_3)^2/2K.
   \ee

   Consequently, it is amusing to note that this stretched
   energy can be rewritten as the light-cone energy due to
   a transverse momentum $p_w$:
   \be
   E_w=\frac{p_w^2}{2P^+}, \qquad
   p_w=w\cdot R_2R_3,
   \label{NEWD}
   \ee
   in which the light-cone mass $P^+=K/R_{11}$! This energy is
   of the form of the energy for an excitation with Kaluza-Klein
   momentum $w$ on a transverse circle with the radius $1/R_2R_3$.
   (After recovering the Planck length, this radius is $l_p^3/R_2R_3$).

   We assert that the spectrum of wrapping membranes is identical
   to that of excitations on a transverse circle. In fact,
   though there exists the apparent degeneration due to the states
   with the same wrapping number $w$ for different combinations
   of $(m,n,s,t)$, however, as noted by Schwarz \cite{SCHRE},
   this degeneration can be easily eliminated by
   considering F-D string spectrum and $SL(2,\integer)$ equivalence.
   Therefore, with wrapping membrane states viewed as
   momentum states, a new flat dimension opens up when
   the compactified 2-torus, on which membranes wrap,
   shrinks to zero (i.e. $R_2R_3\rightarrow 0$). As mentioned
   in the introduction, this is exactly what the IIB/M duality
   requires for a M-theory formulation of IIB string theory.

  \subsection{Membrane Tension}

   To verify that the membrane has the correct tension,
   we note that the light cone energy in M theory
   for a stretched transverse membrane is given by
   \be
    E_m = =\frac{M_w^2}{2P^+},
   \ee
   in which $M_w=wT_{mem}A_{T^2}$.
   $T_{mem}=1/(2\pi)^2$ is the membrane tension, with
   Planck length taken to be unity. $A_{T^2}$ is the
   area of the compactified 2-torus in target space,
   which in our case is regular with sides $2\pi R_2$
   and $2\pi R_3$; so $A_{T^2}=R_2R_3(2\pi)^2$.
   Therefore we have
   \be\label{WEN}
    E_m = w^2R_{11}(R_2R_3)^2/2K.
   \ee
   So in the large-$N$
   limit, we have the equality:
   \be
    E_w = E_m.
   \ee
   This is a perfect and marvelous match! An equivalent
   statement is that our wrapping matrix membrane indeed
   has the correct tension as required by M-theory.

   An astute reader may have noticed that the energy
   (\ref{ENG}) of our membrane configuration contains
   a kinetic part, which also equals $E_w$. We are going
   to clarify its origin in the next section.

\subsection{Adding center-of-mass momentum}

 To better understand the origin of the kinetic part in
 the energy (\ref{ENG}), which is equal to the stretching
 energy $E_w$, we try to add a generic center-of-mass
 momentum to the membrane states (\ref{STT}).

 Let us first examine a closed string. A winding state on
 a compactified circle satisfies
 $x_w(\sigma+2\pi,\tau)=x_w(\sigma,\tau)+2\pi wR$, with $R$
 the radius of the circle. The periodic boundary condition
 is a linear homogenous relation, therefore the momentum
 quantum number can be added as a zero-mode part $x_0$,
 which is linear in $\tau$, such that $x_w + x_0$ still
 satisfies the same equation.

 Now we turn to the orbifold description at large $N$ of
 the same fact, in which $x/R$ becomes the angular part
 of a complex coordinate $z$ and the radius of this circle
 is $NR$. At large $N$ we can make the following
 substitution:
 \be\label{APPX}
  z=NR + i(x_w+x_0)\leftrightarrow NRe^{ix_w/NR}+ix_0.
 \ee
 The equation of motion for the string on the
 $\complex/\integer_N$ orbifold gives the solution
 $x_w/R =w(\sigma-\tau)$, for $w>0$,
 $x_w/R = w(\sigma+\tau)$, for $w<0$ and
 $x_0 = p\tau$. Suppose $w>0$ for definiteness. One
 finds the energy density is given by
 \bea \nonumber
 h &=&|\dot{z}|^2+|z^\prime|^2\\
 &=& (p^2+(Rw)^2-2pRw \cos{\frac{w}{N}(\sigma-\tau)})+(Rw)^2.
 \eea
 It is amusing to observe that in the large-$N$ limit,
 $h$ approaches to $(p-Rw)^2+(Rw)^2$. With $p=0$, the
 kinetic energy is equal to the stretching energy! This
 feature is the same as we have encountered in Eq. (\ref{ENG}),
 indicating that this is a common phenomenon in approaching
 compactification via orbifolding.

 Giving a bit more details, we note that the orbifolding
 solution, in the large-$N$ limit, becomes
 $z\rightarrow NR+i(p-Rw)\tau + iRw\sigma$. Accordingly,
 we should interpret the linear combination $p-Rw$ as the
 momentum in the continuum limit, though naively one might
 have expected that $p$ would be identified as the
 center-of-mass momentum. The same analysis can be applied
 to the continuous membrane wrapping on an orbifold:
 \bea
 \nonumber
  z^2(p,q;\tau) &=& 2^{-1/2}(ik_2\tau
     +NR_2e^{i(N^{-1}(mp+nq)-\omega_2\tau)}),\\
 \label{MWSM}
  z^3(p,q;\tau) &=& 2^{-1/2}(ik_3\tau
     +NR_3e^{i(N^{-1}(sp+tq)-\omega_3\tau)}).
 \eea
 A similar analysis shows again that the momenta in the continuum
 limit are shifted by the wrapping number: $p_a = k_a - NR_a\omega_a$,
 for $a=2,3$. Thus with $k_a=0$, a membrane wrapping on a torus is not
 at rest!

 The states in Eqs.~(\ref{APPX}) and (\ref{MWSM}) can be generalized
 to matrix membranes, with center-of-mass momentum added:
 \bea
 \label{STT1}
 \nonumber
  Z^2 &=& 2^{-1/2}(iR_{11}k_2\tau/2\pi K
     + c_2(\tau) U^{m} V^{n})\otimes \hat{V}_2, \\
  Z^3 &=& 2^{-1/2}(iR_{11}k_3\tau/2\pi K
     + c_3(\tau) U^{s} V^{t})\otimes \hat{V}_3
 \eea
 where, for a regular torus in target space, we have set $Z^1=0$.
 As in the previous section, we take $c_{a} = c_{a0}e^{-i\omega_{a}\tau}$
 with $\omega_{a}$ $(a=2,3)$ in Eq.~(\ref{OMEG1}). In large-$N$ limit,
 the matrix membrane configuration in Eq.~(\ref{STT1}) is expected to be
 equivalent to the continuous configuration in Eq.~(\ref{MWSM}) up to
 a normalization.

 The kinetic energy density in the discrete setting,
 $T:=R_{11}^{-1}|\dot{Z}_a|^2$, can be easily computed:
 \be\label{KED}
  T = \frac{1}{2R_{11}} [(\frac {R_{11}k_a}{2\pi K})^2
   +\omega_a^2|c_a|^2 -((\frac{R_{11}k_a}{2\pi K})
    \omega_ac_a U^{m_a}V^{n_a} + h.c.)]\otimes \unit_{N^2},
 \ee
 in which the summation is over $a=2,3$, $m_2=m$, $n_2=n$,
 $m_3=s$, $n_3=t$. Adding the $T$ in (\ref{KED}) to the $V$
 in Eq.~(\ref{ENGD}), we get the total energy density,
 which results in the total energy of a wrapping membrane
 configuration with momenta and wrapping:
 \be\label{Energy}
  E_{k_2,k_3;w} = {R_{11}\over 2 K}(k_a-\frac
 {2\pi K}{ R_{11}} \omega_a |c_a|)^2+ E_w.
 \ee
 We are glad, just for convenience, to rewrite
 \bea
  \omega_2 &=& (\sqrt{2}\pi)^{-1}NR_{11}R_3\sin{(\pi w/KN^2)}, \\
  \omega_3 &=& (\sqrt{2}\pi)^{-1}NR_{11}R_2\sin{(\pi w/KN^2)}, \\
  |c_a| &=& NR_a/2\pi.
 \eea
 Accordingly, the genuine momentum is identified to be
 \be
  p_a=k_a - (\sqrt{2}\pi)^{-1} KN^2R_2R_3 \sin{(\pi w/KN^2)}
  \rightarrow k_a - R_2R_3 w/\sqrt{2}.
 \ee

 In closed string theory, the quantization of momentum is a
 consequence of the single-valuedness of the translation
 operator $e^{i2\pi R_a\hat{p}_a}$ (no summation on $a$).
 In our present case, we can also impose the quantization
 condition at large $N$; as a result $p_a = l_a/R_a$ where
 $l_a$ is an integer (or equivalently, one has
 $k_a = l_a/R_a  + R_2R_3 w/\sqrt{2}$). Note that the
 domain of $(p_2,p_3)$ is just the lattice dual to the
 target torus; therefore, the above equation is once more
 a manifestation of the covariance for $E_{k_2,k_3;w}$
 under the $SL(2,\integer)$ transformations over the
 target torus.

 In summary, our explicit construction of wrapping membrane
 states, the appearance of $SL(2,\integer)$ symmetry in
 their spectrum, and an emergent flat dimension as well,
 all these combined together, constitute strong evidence
 for our quiver matrix mechanical model to be a non-perturbative
 formulation of IIB string theory, which naturally exhibits
 IIB/M duality.

 \section{Discussions and Perspectives}
 \label{SEC6}

   Problems that remain for future study and some perspectives
   are collected in this section. Some of them are just
   slightly touched in this work. (As mentioned before, we
   will leave to a sequential paper the discussions on generic
   moduli for the VEV of $Z^a$ ($a=1,2,3$), which would allow
   both the SYM world-volume torus and the compactified
   target space torus non-regular, and would explicitly
   demonstrate the $SL(2,\integer)$ duality of our approach
   to IIB string theory.)

   \begin{enumerate}

   \item There have been three different ideas to deal with
   IIB/M duality, i.e. M-theory on a 2-torus should be dual
   to IIB string theory on a circle: namely the wrapping membrane
   states, the vector-scalar duality in three dimensions and the
   magnetic charges of membranes, respectively (see for example
   \cite{SS}). We did not explore the latter two in the present
   work. In fact, a naive definition of magnetic charge such as
   $Tr[Z^a,Z^{a^\prime}]$ for finite matrix configurations
   vanishes identically.

   \item For IIA/M duality, there is a complete dictionary
   of the correspondence between the spectrum, as well as
   operators, between IIA D0 brane and M-theory objects, in
   the Matrix String Theory {\it a la} DVV \cite{DVV}. As for IIB/M
   duality a comprehensive dictionary for spectrum and
   operators between the two sides remains to be worked out.
   Moreover, in this work we recovered only part of
   the $SL(2,\integer)$-invariant spectrum. A more detailed
   study to demonstrate the full $SL(2,\integer)$ symmetry
   is, in principle, possible in the present framework for
   IIB strings, and we leave it for future research. How
   to relate this approach to other nonperturbative IIB
   string theory, such as IIB matrix model \cite{IKKT}
   or that based on the D-string action \cite{HW},
   is another interesting issue to address.

   \item A discrete approach, the string bit model, has been
   proposed to IIB string theory before \cite{BT95}. The
   difference between our quiver mechanics approach and the
   string bit model is the latter discretizes the non-linear
   sigma model for string theory on sites of a lattice, while
   we deconstruct SYM with part of matter fields living on
   links.  Recently the string bit model gained revived
   interests \cite{Ver02} in the context of the BMN
   correspondence \cite{BMN}. On the other hand, BMN have
   devised a (massive) matrix model in PP-wave background.
   How to do orbifolding with this M(atrix) model remains
   a challenge.

   \item Probing spacetime with strings has revealed T-duality
   of spacetime; i.e. stringy dynamics gives rise to new
   features of spacetime geometry. In the same spirit, one expects
   that probing spacetime with membranes would expose new, subtle
   properties or structures in spacetime geometry too. Actually
   it has been suggested \cite{FHRS} that M(atrix) Theory
   compactified on circle, on two- and three-torus are tightly
   related to each other. We leave the exploration in the
   framework of quiver mechanical deconstruction to the future.

   \item The appearance of non-unitary fields living on links is
   a generic phenomenon in discrete models. In addition to our
   previous work \cite{DS}, a few other authors also paid attention
   to the effects of non-unitary link fields \cite{LM,LMP,JLM}.
   Here we would like to emphasize the significant role of
   non-unitary link fields in dynamics of geometry, which is
   worthwhile to explore in depth.

   \end{enumerate}

 {\bf Acknowledgement}

  YSW thanks the Interdisplinary Centor for Theoretical
  Sciences, Chinese Academy of Sciences (Beijing, China)
  for warm hospitality during his visit, when the work
  was at the beginning stage. JD thanks the High Energy
  Astrophysics Institute and Department of Physics, the
  University of Utah, and Profs. K. Becker, C. DeTar and
  D. Kieda for warm hospitality and financial support;
  he also appreciates the helpful discussion with B. Feng
  and K. Becker.

 \end{document}